\documentclass[12pt]{aastex62}
\newcommand{\mytilde}{\raise.17ex\hbox{$\scriptstyle\mathtt{‌​\sim}$}}

\newcommand{\Ef}{ {\mathcal E} }

\newcommand{\vE}{\mbox{\boldmath $\mathcal E$}}         

\def\mw{{microwave}}
\def\Mw{{Microwave}}
\def\kw{{Konus-\textit{Wind}}}
\def\hsi{{\textit{RHESSI}}}
\usepackage[euler]{textgreek}

\ifx \doiurl    \undefined \def
\doiurl#1{\href{http://dx.doi.org/#1}{\textsf{DOI}}}\fi \ifx
\adsurl    \undefined \def
\adsurl#1{\href{http://adsabs.harvard.edu/abs/#1}{\textsf{ADS}}}\fi
\ifx \arxivurl  \undefined \def
\arxivurl#1{\href{http://arxiv.org/abs/#1}{\textsf{arXiv}}}\fi

\usepackage{graphicx,natbib,color,amsmath,subfigure}
\usepackage{epsfig}          
\usepackage{graphicx}        
\usepackage{epstopdf}
\usepackage{amssymb}        
\usepackage{xcolor}
\usepackage{ulem}

\received{ 2019}
\revised{ 2019}
\accepted{\today}
\submitjournal{ApJ}

\shorttitle{Electron Acceleration Process}
\shortauthors{Altyntsev et al.}

\begin{document}

\title{Rapid Variability in the SOL2011-08-04 Flare: Implications for Electron Acceleration}

\correspondingauthor{Nataliia Meshalkina}
\email{nata@iszf.irk.ru}

\author{Alexander T. Altyntsev}
\affil{Institute of Solar-Terrestrial Physics\\
126a Lermontov st.\\
Irkutsk, 664033, Russia}
\author{Nataliia S. Meshalkina}
\affiliation{Institute of Solar-Terrestrial Physics\\
126a Lermontov st.\\
Irkutsk, 664033, Russia}

\author{Alexandra L. Lysenko}
\affiliation{Ioffe Institute, Polytekhnicheskaya, 26, St. Petersburg, 194021 -  Russia}

\author{Gregory D. Fleishman}
\affiliation{New Jersey Institute of Technology, University Heights, Newark, NJ 07102-1982 -  United States}

\begin{abstract}

Particle acceleration in solar flares remains an outstanding problem in solar physics. It is yet unclear which of the acceleration mechanisms dominates and how exactly is the excessive magnetic energy transferred to the nonthermal and other forms of energy. We emphasize, that the ultimate acceleration mechanism must be capable of efficiently working in the most extreme conditions, such as the shortest detected time scales and the highest acceleration efficiency. Here we focus on detailed multiwavelength analysis of a very initial phase of the SOL2011-08-04 flare, which demonstrated prominent short subpeaks of nonthermal emission during filament eruption associated with the flare. We demonstrate that the three-dimensional configuration of the flare, combined with timing and spectral behavior of the rapidly varying component, put very stringent constraints on the acceleration regime. Specifically, the rapid subpeaks are generated by short injections of nonthermal electrons with a reasonably hard, single power-law spectrum and a relatively narrow spread of pitch-angles along the mean magnetic field. The acceleration site is a compact volume located near the top of extended coronal loop(s). The electrons are
accelerated up to several hundreds of keV promptly, with the characteristic acceleration time  shorter than 50\,ms. We show, that these properties are difficult to reconcile with widely adopted stochastic acceleration models, while the data inescapably require acceleration by a super-Dreicer electric field, whether regular or random.

\end{abstract}

\keywords{Radio Emission, Solar; Polarization; X-Ray Bursts, Flares}

\section{Introduction}

Solar flares are, perhaps, the most powerful accelerators of charged particles in the solar system.
However, in spite of extensive observational, theoretical, and modeling efforts over dozens of years,
we are still far away from detailed understanding of acceleration mechanisms and clear scenario of
how exactly the excessive energy stored in nonpotential coronal magnetic field is converted during flares
to other forms of energy, including nonthermal energy. In particular, the relationship between
various time scales, ranging from very short, subsecond, to very long, multi-hour, is unclear.
We do not know yet if subsecond episodes of particle acceleration are produced by the same
or different acceleration mechanism as compared with a more sustained acceleration that lasts
from minutes to hours, if they occur in the same or different volumes,
if acceleration primarily occurs during these short episodes, while a more gradual component
is a cumulative effect of many such episodes.

Acceleration episodes manifest themselves as distinct short subpeaks
of HXR emission which were proposed to represent a signature of direct
precipitation of non-thermal electrons into the solar atmosphere
dense layers. Those electrons  were believed to be accelerated in
discrete reconnection episodes in coronal structures of magnetic
field \citep{1983ApJ...271..376K}. It was suggested
that duration of an elementary burst was controlled by the size of
interacting current filaments \citep{1984SoPh...94..341S,
1993ApJ...418..912L}, or by typical time of reconnection processes
\citep{1996SoPh..167..321L}. Therefore, the flare configuration and time-of-flight relationships provide a basis for considering the acceleration process within the frameworks of either a ``small-scale'' model, which
assumes that the electrons are accelerated
promptly high in the corona and then precipitate down with constant velocities, or a ``large-scale'' model, which assumes that the electron acceleration is distributed over a large volume such as a significant portion of a flaring loop \citep{1996AIPC..374..300A}.

Whether within the small- or large-scale model, the 
mechanisms of particle acceleration in solar flares are still under debate and
development. The list of proposed acceleration mechanisms is extensive:
acceleration by DC electric fields in the coronal loops or in current sheets due to
reconnection \citep{1994ApJ...435..469B, 2004ApJ...604..884Z,
2016ApJ...827L...3V, 2019PPCF...61a4020V}, betatron
acceleration in collapsing magnetic loops \citep{1997ApJ...485..859S,
 2012A&A...546A..85G}, acceleration in kink-unstable twisted magnetic loops \citep{2014A&A...561A..72G},
 acceleration by Alfven waves in loops
\citep{2008ApJ...675.1645F}, by shock waves generated by plasma outflows
from current sheet \citep{1997ApJ...485..859S,
2006A&A...454..969M, 2013PhRvL.110e1101N}, resonant stochastic acceleration in small-scale fields of plasma
turbulence \citep{1996ApJ...461..445M, 2004ApJ...610..550P}, acceleration by helical
turbulence in twisted magnetic field \citep{2013MNRAS.429.2515F}, and
non-resonant acceleration by strong large-scale turbulence
\citep{2009ApJ...692L..45B}.

HXR data provide the primary diagnostics of electron acceleration in flares,
but it has severe limitations and biases. With a typical HXR spectrum falling
with energy as a power-law with reasonably large indices, $\gamma=3-8$,
the numbers of detected photons falls off quickly with energy such as
it is  impossible to either image the HXR sources or
isolate short temporal HXR subpeaks from the noise above $\sim100$~keV in a typical flare.
On top of that, the HXR emission intensity is weighted with the number density
of the ambient thermal plasma; thus, the HXR images are typically dominated by chromospheric footpoints,
where a so-called thick-target HXR emission is produced \citep[e.g.,][]{2011SSRv..159...19F}.
In rare cases, the coronal portion of a flaring loop can be sufficiently dense
(of the order of $10^{11}$cm$^{-3}$ or above) to render the loop collisionally thick for the accelerated electron.
In such cases, coronal thick-target sources can be dominant \citep{2012ApJ...755...32G},
while the precipitation to the footpoints suppressed \citep[see, however][]{2018ApJ...867...82D}.

A significant source of data that nicely complements the HXR
diagnostics of the dynamics of
electron acceleration comes from microwave observations, whose sensitivity
to emission of nonthermal electrons in coronal magnetic structures is
considerably higher \citep{2012ApJ...758..138A} compared with HXRs.
Combination of spectral and imaging information in both HXR and
microwave domains has already permitted to pinpoint the flare energy
release/electron acceleration regions in a few case studies
\citep{2004SoPh..221...85M, 2011ApJ...731L..19F,
2013ApJ...768..190F, 2016ApJ...826...38F}.

It is important to realise that unlike to HXR emission, which
is produced by a single well--understood emission mechanism,
bremsstrahlung, the radio emission can be produced by a rich variety of
incoherent or coherent processes. This is especially true for short time
structures, which often have a rather narrow instantaneous spectral
bandwidth \citep{1969resp.book.....Z, 1986SoPh..104...99B,
2007SoPh..242..111A}. Even  a relatively tiny population of nonthermal electrons
can produce such bursts by a coherent emission mechanism.
On the contrary, broadband microwave background bursts,
even having a rather short, subsecond duration, are believed
to be generated by an incoherent gyrosynchrotron emission mechanism with,
perhaps, a contribution from the free-free process
 \citep[e.g.,][]{2008ApJ...677.1367A, 2018ApJ...867...84G}. Such gyrosynchrotron bursts are known
to be extremely helpful in probing the magnetic field and non--thermal
electron distribution \citep{2013SoPh..288..549G, 2018ApJ...863...83G}.

The microwave continuum bursts last typically from several seconds  to minutes
\citep[see, for example,][]{2004ApJ...605..528N, 2008ApJ...677.1367A, 2008ApJ...684.1433F, 2018ApJ...856..111L},
while those with a subsecond duration are detected
more rarely \citep{1980A&A....87...58K, 2000SoPh..195..401A, 2004SoPh..221...85M,
 2016ApJ...822...71F, 2018ApJ...867...84G}, likely, because the latter requires that some
special conditions are met: (i) short and well separated episodes of electron acceleration; (ii) a high acceleration  rate of electrons up to relativistic
energies; and (iii) inefficient trapping of these electrons  in the flaring coronal loops. However, when such subsecond subbursts are detected, their study can substantially narrow the list of possible acceleration processes given that only a subset of those processes will turn out capable of supporting that extreme acceleration efficiency at that short time scales.

It is intuitively clear that the acceleration processes can
be more conclusively studied in flares with a
relatively simple configuration of flare sources, in which no or only
a weak trapping of accelerated electrons occurs and, thus, emission
from the acceleration site directly dominates
 \citep[see, e.g.,][]{2011ApJ...731L..19F, 2016ApJ...826...38F}.

In this work we study the primary energy
release in the initial phase of the 2011 August 4 flare.
The 2011 August 4 flare of GOES class M9.3 occurred in the central part
of solar disk N17W37 (NOAA 11261) and reached its maximum at
03:56:30 UT. During a filament eruption prior to the flare impulsive
phase, a unique sequence of broadband pulses was observed.
 We concentrate on analysis of a few episodes of short
subsecond subbursts, both narrowband ones recorded in the radio domain
only, and broadband subbursts recorded in both radio and X-ray domains, as
compared with a more gradual underlying burst. We are
specifically interested in characterizing spatial relationships between
the sources of the short subsecond subbursts and source(s) of a more
gradual emission and quantifying the sources in terms of their physical
parameters (sizes, magnetic fields, number densities, etc) with the
goal of narrowing down the list of possible acceleration mechanisms.

\section{Observations}

As has been explained in the Introduction, we need to characterize a
flare with as high spatial and
temporal resolution as possible. This requirement motivates our
selection of two imaging radio instruments working at three different
microwave frequencies, and three different X-ray instruments, which
collectively provide high spatial, spectral, and temporal resolution of
the X-ray emission. This section gives an outline of the instruments
and the data used in our study.

\subsection{Instrumentation}
\label{S_Instr}

\subsubsection{\Mw\ Domain}
\label{S_Instr_mw}

The \textit{Siberian Solar Radio Telescope} (SSRT) records one-dimensional (1D) radio
images (scans) at 5.7~GHz with 14~ms cadence using East-West (EW) and North-South (NS) antenna arms \citep{2003A&A...400..337A, 2012SoPh..280..537M}. During
the observations, the antenna beam width at the level of full width at half
maximum (FWHM) was 18 and 20 arcsec in the EW and NS directions, respectively. 
We employed fast SSRT light curves to identify short subpeaks with the duration of less than or about a second. Fast 1D scans were used to study locations of the subpeak sources. SSRT also produces two-dimensional (2D) images of solar disk every 2-3 minutes \citep{2003SoPh..216..239G, 2013PASJ...65S..19K}. 2D images were used to study the morphology and location of the background \mw\ burst source.

The \textit{Nobeyama Radio Heliograph} \citep[NoRH,][]{1994IEEEP..82..705N} produces full-sun 2D images  at
17 GHz (intensity and polarization) and 34 GHz (intensity only) with the 0.1~s cadence. The
NoRH spatial resolution was 13 arcsec at 17 GHz and 8 arcsec at 34~GHz at the time of the flare under study.
The imaging was employed to study locations and morphology of the flare at high frequencies.
To study the fast source dynamics we used imaging with the highest possible 0.1~s cadence.

The \textit{Nobeyama Radio Polarimeters} measure the total power (TP) radio emission at six \mw\ frequencies from 1--35~GHz  with the cadence
of 0.1 s \citep[NoRP,][]{1979PRIAN..26..129T}. We used the NoRP data to study the evolving \mw\ spectrum. However, given a rather week signal at 35~GHz, we also employed the cross-correlation curves of the longest NoRH baselines to recover evolving radio flux of the burst at 34~GHz.

\subsubsection{X-ray Domain}

The \textit{Reuven Ramaty High-Energy Solar Spectroscopic Imager} \citep[\hsi,][]{2002SoPh..210....3L} records individual photons with high spectral resolution between 3~keV and 17~MeV with nine front and nine rear detectors.  \hsi\
employs rotational synthesis imaging technique possible due to spacecraft rotation with 4~s; thus,
higher time resolution light curves are not easily available.
We employed the \hsi\ data to produce X-ray images and spectra of the sources of the smooth emission component, while used two other instruments to study fine temporal structure of X-ray emission.

The \textit{Fermi Gamma-Ray Burst Monitor} \citep[GBM,][]{2009ApJ...702..791M} is one of
instruments onboard the \textit{Fermi Space Telescope} \citep{2009ApJ...697.1071A}, which provides the light curves {in the spectral range from $\sim8$ keV to $\sim40$ MeV} with temporal resolution of 256/64 ms, while the GBM spectra were available with the cadence of 1.024~s. The detector area is 128 cm$^{2}$.

The Spectrometer KONUS on board of the \textit{Wind} spacecraft \citep[\kw,][]{1995SSRv...71..265A} works in either waiting or triggered mode. The detector area is $\approx100$~cm$^{2}$.
In the triggered mode, employed  here, the light curves are recorded in three energy bands (G1 18--75~keV, G2 75--308~keV, and G3 308--1050~keV in the event under study) with variable time resolution from 2 to 256 ms. Multi-channel spectra are taken over uneven time intervals, whose duration is determined adaptively depending on the count rate and varies between 64~ms and $\sim$8~s. {We employed the time domain \kw\ data to study variations of the HXR spectral hardness with a subsecond time resolution.}

\subsubsection{Context Data}

The \textit{Atmospheric Imaging Assembly} \citep[AIA:][]{2012SoPh..275...17L} onboard the \textit{Solar Dynamics Observatory}
(\textit{SDO}) is a narrowband UV imager that records full-disk solar UV/EUV emission in 10 narrow passbands with 1.5\arcsec\ angular resolution and 12~s cadence.
We employed the AIA imaging data in a few ``coronal'' passbands sensitive to hot plasma to investigate the flare morphology.

The \textit{Helioseismic and Magnetic Imager} \citep[HMI:][]{2012SoPh..275..207S} onboard the \textit{SDO} provides full-disk solar magnetograms, dopplergrams, and white-light images with a subarcsecond spatial resolution and 45~s to 12~minute time cadence depending on the data product. In this study, we employed full-disk vector magnetograms to initiate 3D modeling of the flare.

\subsubsection{Analysis and Modeling Tools}

In most cases, standard SSW analysis tools were used. In particular, the RHESSI data were analyzed
with the OSPEX and the imaging software developed by the RHESSI team \citep{2002SoPh..210....3L}
which permits obtaining spectra, full-disk light curves, and images at various energy bands, and perform spectral fits.

GX Simulator \citep{2015ApJ...799..236N, 2018ApJ...853...66N} is an interactive tool (an IDL-based widget freely distributed via SSW) developed for 3D modeling of active regions and solar flares. It supplies the user with a possibility of automated production of 3D magneto-plasma data cube built for a user-defined field of view and spatial resolution. The 3D magnetic model is created by nonlinear force-free field reconstruction \citep{2017ApJ...839...30F} initiated with the corresponding photospheric vector magnetogram automatically downloaded from the HMI/SDO site. We employed the GX Simulator for 3D modeling of the flaring loops and computation of the associated electromagnetic emission.

\subsection{X-ray and microwave light curves}
\label{S_Obs_LC}

\begin{figure}    

  \centerline{\includegraphics[width=0.8\textwidth,clip=]{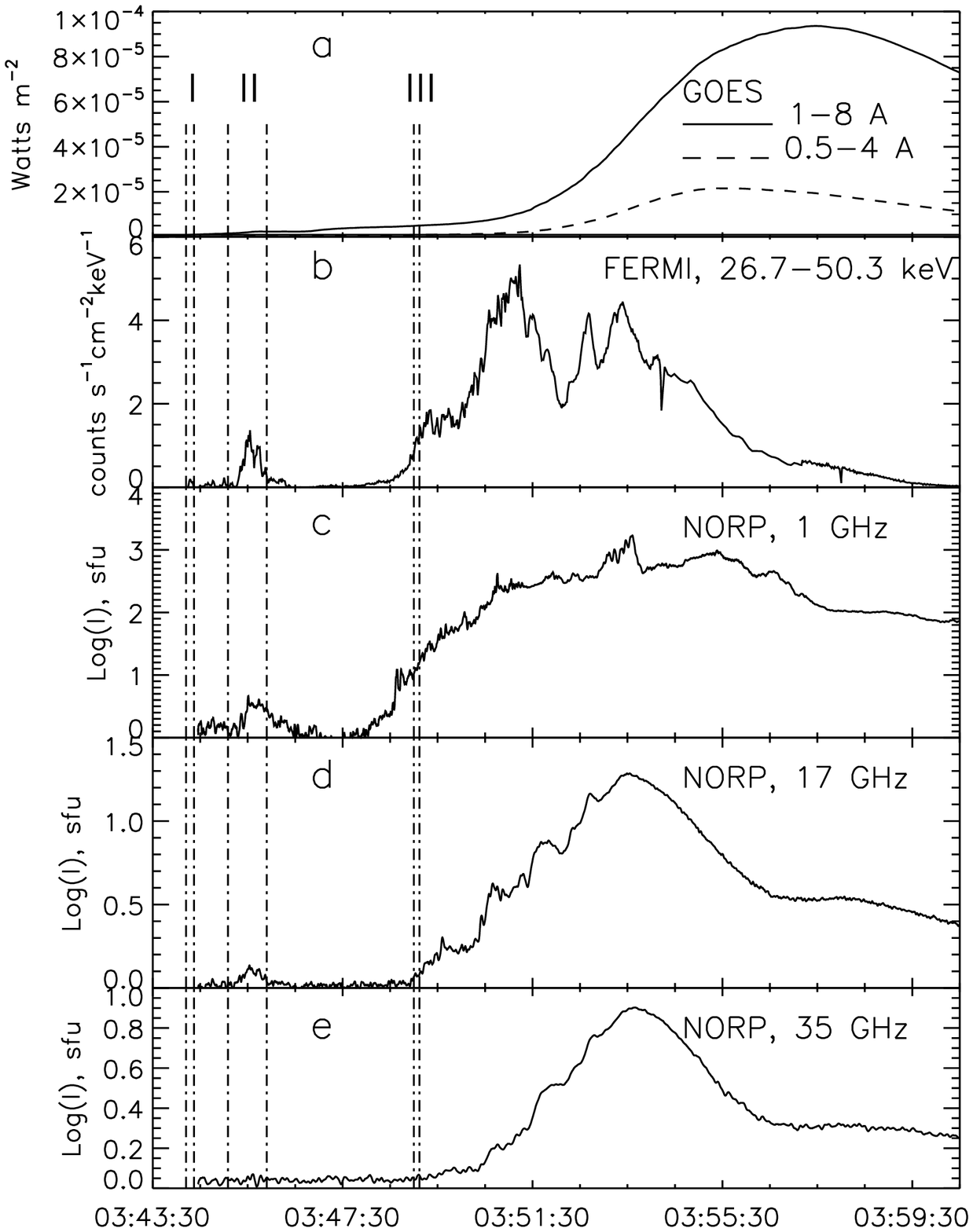}
             }
             \caption{The flare light curves of X--ray (a,b) and microwave emission (c--e).
             Three intervals with subpeaks of about a second length are marked by vertical lines.
                     }
  \label{F1-simple}
  \end{figure}

Figure~\ref{F1-simple} displays a representative subset of the
X-ray and microwave light curves.  A few series of short subpeaks
with duration about a second are seen  at the initial
stage of the flare. To highlight  weak emission enhancements before the
main flare phase, we use the log scale for the radio light curves. The time ranges of the fast variations
are marked by the vertical lines in Figure~\ref{F1-simple}:  group I
(03:44:15 - 03:44:22 UT),  group II (03:45:00 - 03:45:54 UT) and
group III (03:49:00 - 03:49:08 UT). The time profiles of the smoothed (background) microwave
emission are overall similar to each other in a rather broad range
of frequencies, 1 -- 35 GHz, which favors a single incoherent
mechanism of the background emission. The soft X-ray emission raised slowly at the initial stage of the flare.

Groups I and II occurred a few minutes before the flare
impulsive phase. During the first group there were subpeaks with a
"one-to-one match" correspondence between pulses at 5.7 and 17 GHz,
and HXR emission pulses. The NoRP data with the time resolution of 0.1 s
are not available and this group was too weak for quantitative
analysis. More detailed and reliable data are available for the
second group of the subpeaks. The subpeaks of the second group were reasonably
strong as they occurred during the microwave burst with flux density
reaching hundred sfu at 3.75---9.4~GHz.

The third group occurred later, at the beginning of the flare impulsive phase, while at 5.7 GHz only. Several oscillations of emission flux with approximately 0.5~s period occurred during 8 seconds. Unlike previous intervals with the subpeaks on NoRP and HXR light curves, no counterpart to the oscillations at 5.7 GHz was observed at either NoRP or
HXR light curves. We conclude that in this case, the subsecond pulses are narrow-band and,
thus, are likely generated by a coherent mechanism.

\begin{figure}

 \centerline{\includegraphics[width=0.8\textwidth,clip=]{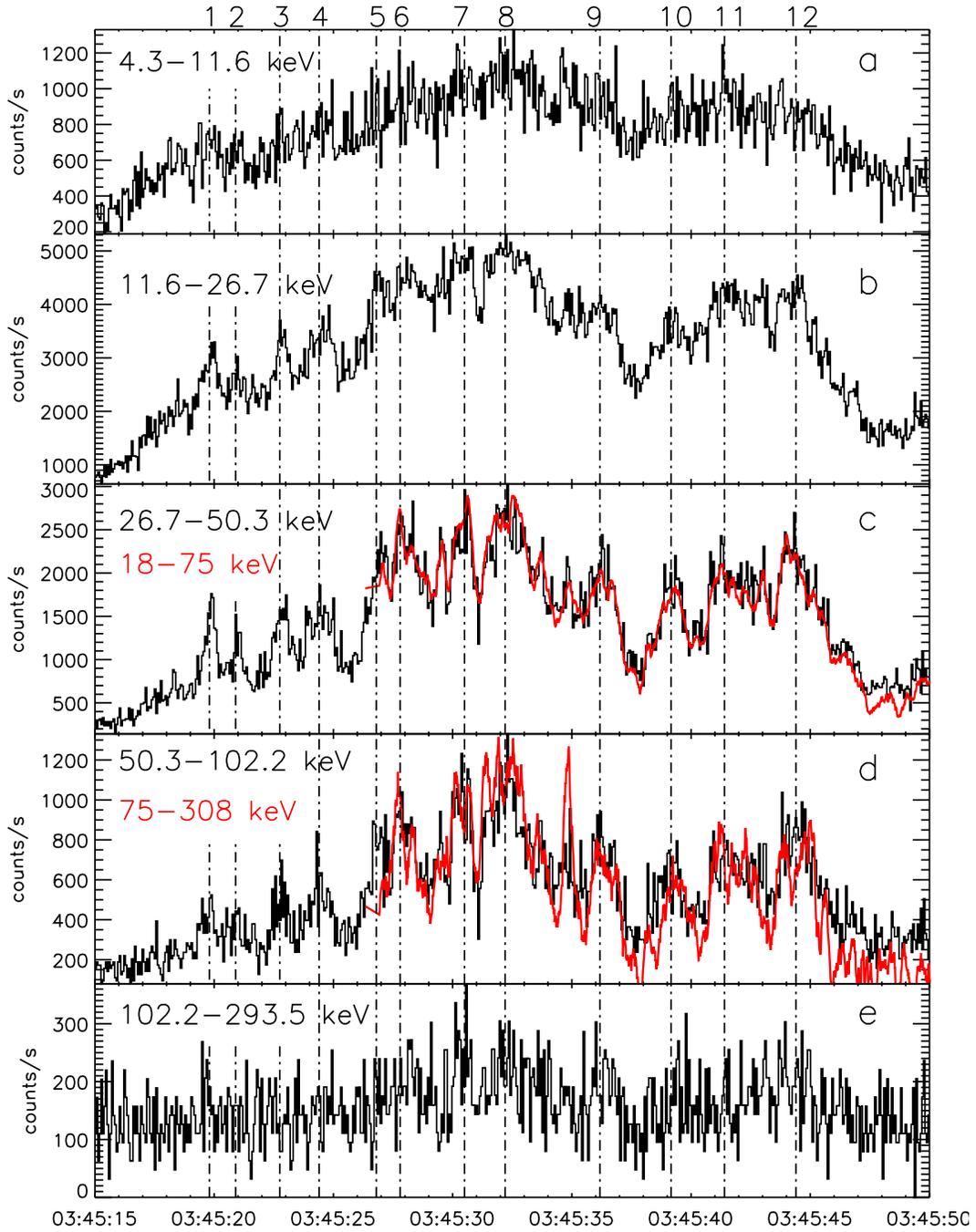}
             }
              \caption{Interval II. Intensity of HXR emission in
             channels GBM and \kw\ 18--75 and 75--308~keV
             (red curves). \kw\ data in fast mode are available after 03:45:27.5 UT. Rate in counts/s.
             Vertical lines mark the prominent subpeaks.
                     }
  \label{F2-simple}
  \end{figure}

\begin{figure}

  \centerline{\includegraphics[width=0.8\textwidth,clip=]{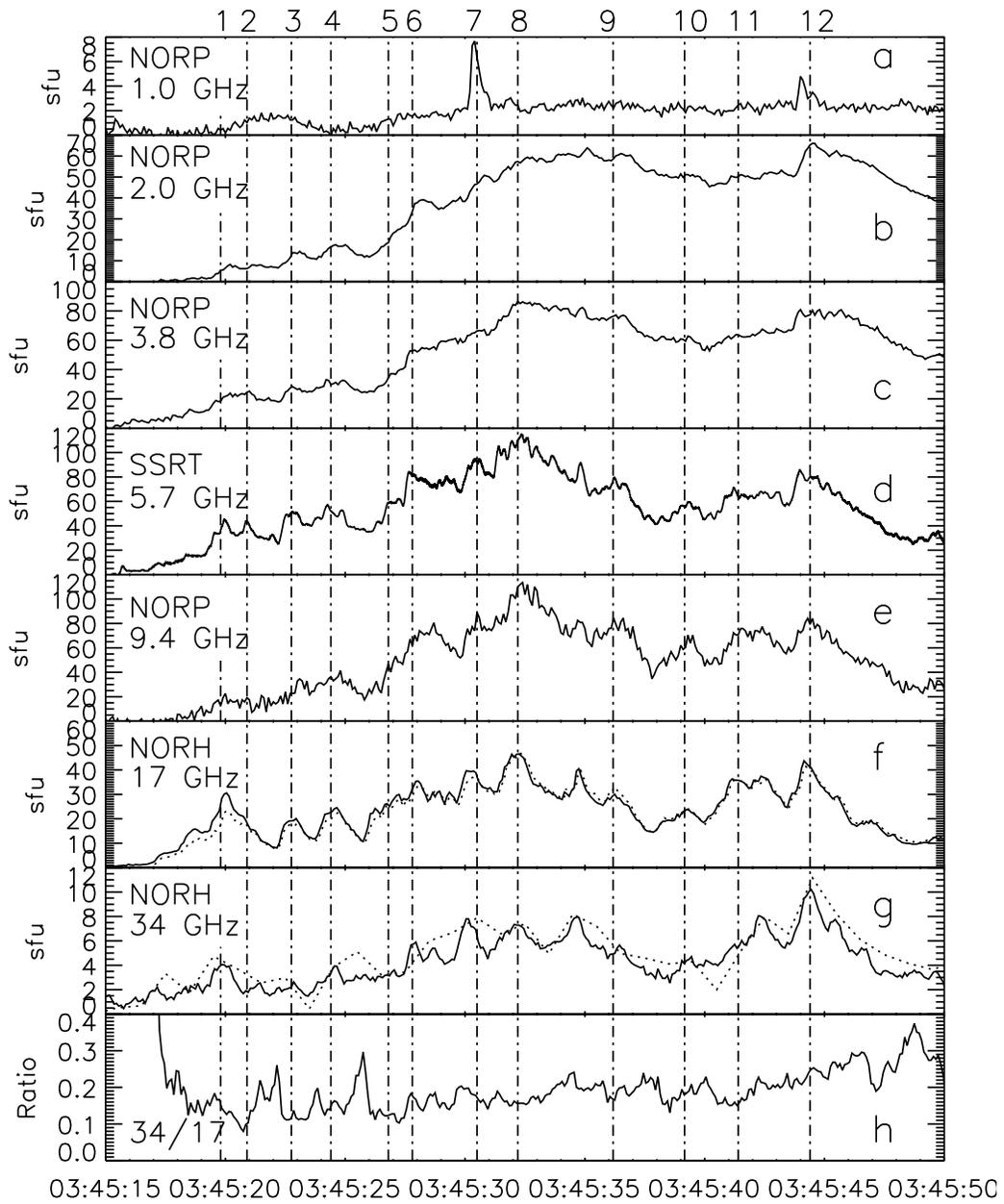}
             }
             \caption{Interval II. Time profiles of microwave emission.
             Data from the NORP (1.0--3.75 and 9.4 GHz),
              SSRT (5.7 GHz) and NORH (17 and 34~GHz).
             At 17 GHz and 34 GHz the dotted curves are fluxes and the solid curves are the correlation plots. h) ratio of the sfu-normalized correlation curves at 34 and 17 GHz.
                                  }
  \label{F3-simple}
  \end{figure}

Let us study the second group in more details
(Figures~\ref{F2-simple}, \ref{F3-simple}). There is a
"one-to-one" correspondence between individual peaks in the HXR
and microwave emission. The subpeak durations (the time intervals
between signal minima) are within the range 0.7--1.8 s.
Figure~\ref{F2-simple} presents the GBM-Fermi light curves recorded with 64~ms bins.
The calculated errors are in range from 16 to 36 counts/s in the GBM channels. The reliability of the short pulses in hard X-rays is confirmed by the comparison of time profiles recorded by the GBM and \kw. The \kw\ data in the
triggered mode\footnote{Data for all flares registered by \kw\ in the triggered mode are available online via \url{http://www.ioffe.ru/LEA/kwsun/}} are available after 03:45:27.5 UT. The \kw\ profiles (Figure~\ref{F2-simple}c,d, dotted curves) are smoothed with a window width of 256 ms, and the amplitudes are normalized to the amplitudes of the GBM profiles. 

The hard X-ray subpeaks stand out cleanly at two energy channels: 26.7--50.3~kev and 50.3--102.2~kev where their amplitudes are comparable with that of the more gradual background component. To aid the eye to see the correlation between various light curves,
the vertical lines mark times of the most prominent
subpeaks with duration about 1~s in the GBM 26.7--50.3 profile (Figure~\ref{F2-simple}c).
In this channel the GBM rate are sufficiently large ($ > 1000$ counts/s) to separate the subpeaks.
At lower energy channels the subpeaks are less pronounced. At 102.2--293.5~kev the count rate is lower and the peak correspondence is apparent for large peaks only. Reliability of the subpeaks is confirmed by the \kw~records. There is a clear correspondence between the profiles in Figure~\ref{F2-simple}c. Correlation is lower in Figure~\ref{F2-simple}d in energy channels G2 (75--308~keV), while  channel G3 is not shown because there is no significant signal above the level of statistical fluctuations.

The microwave light curves in Figure~\ref{F3-simple} were obtained with the NoRP at 1.0, 2.0, 3.75 and 9.4~GHz, and by the SSRT at 5.7 GHz, while the NoRP data at 17 and 35~GHz were too noisy. For that reason, we employed the NoRH data at 17 and 34~GHz. The dotted curves in panels (f) and (g) are obtained by integrating the sequences of NoRH images over the field of view. This technique worked well with the original 0.1~s cadence at 17~GHz, while  did not work at 34~GHz due to significantly lower signal-to-noise ratio.  Thus, in panel (g), the dotted curve shows the radio flux obtained with 1 sec cadence and then smoothed with 3 s cadence, which is insufficient to resolve individual subpeaks. To recover the subpeaks at 34~GHz, we employed the NoRH correlation plots involving only the longest baselines, which are only sensitive to the most compact sources of emission. The validity of this approach is confirmed by 17~GHz data, where the profiles of the normalized correlation plots (solid) and the fluxes (dotted) closely match each other; thus, the correlation plot can be straightforwardly renormalized to represent the burst flux in sfu. The corresponding solid curve in panel (g) also matches the dashed one on average, while recovering the fine temporal structure at 34~GHz. Thus, the noise levels do not exceed 5~percent at all frequencies except 34~GHz, where it is two times larger. As in the HXR case, the profiles become smoother at the lower frequencies. Note, however, prominent short, narrowband subpeaks at 1 GHz numbered 7 and 12, which are likely produced by a coherent mechanism. 
The observed microwave spectrum is typical for gyrosynchrotron emission with a spectral peak within the 5.7--9.4 GHz range \citep{2004ApJ...605..528N}.
Thus, the microwaves and X-ray light curves
at the frequencies 5.7--17 GHz and energies 26.7--102.2~keV are similar to each other at all time scales, including the subsecond structure. 

Although the light curves at various frequencies/energies are highly correlated,
they show measurable delays between each other.
To quantify the time lags, all
light curves,  during the time interval of 03:45:15--03:45:55 UT, were interpolated down to 14 ms cadence to match the best available time resolution provided by SSRT data.
From the resulting time profiles, we subtracted the running mean with 2.1~s smoothing window and computed the lag correlation between the light curves (see Table~\ref{T-simple}). 
The correlation coefficient and the delay values  depend on the width of the smoothing window only slightly. Within  smoothing windows of 1.3--2.8 s, the calculated delays are within the error limits in Table~\ref{T-simple}. The last column contains estimates of the ``weighted'' energy of emitting electrons. The obtained correlation coefficients are high except the 4.3--11.6~keV channel. In all cases, the calculated delays were significantly shorter than the subpeak duration. In HXR domain, the subpeaks at low energies
occur a few dozens of milliseconds later than those at the high energies.
The microwave light curve at 5.7~GHz is delayed  by roughly 70~ms, while those at 17~GHz and 34~GHz  by roughly 120~ms relative to the 102.2--293.5~keV channel.

The subpeak durations in the photon energy range
of 26.7--293.5~keV is within 0.7--1.8~s, while the rise and decay times are comparable to each other. This property alone tells us that the electron transport in this event necessarily includes precipitation to the
chromospheric footpoints. Indeed,
such a rapid decay of the HXR emission requires accordingly rapid
collisional losses of the nonthermal electrons with energy of several hundred keV, which can only happen in a dense
chromospheric footpoint plasma, rather than in the coronal portion of the flaring loops.  Specifically, to ensure the collisional loss time of less than 0.5~s,
the plasma density should exceed $2\times10^{12}$cm$^{-3}$ \citep{1965RvPP....1..105T}, which is unavailable in the coronal portion of the flaring loops.

\subsection{Configuration of the sources}
\label{S_Obs_Conf}

Spatial structure of the sources, their relationships, and sizes are
of primary importance to quantify electron acceleration in flares.
Overall contextual information about the flare morphology and, to
some extent, its dynamics, can be obtained from EUV data.
\citet{2014ApJ...785...88Z} investigated  evolution of this active region, NOAA 11261,
during several days prior to the flare. There were three leading
spots of the south polarity and a tail plage with the northern
magnetic field. Two days before the flare under study, a filament
appeared  in 193 \AA \ and 304 \AA \ passbands along the neutral
line. It destabilized on August 3, after M6.0 flare at 13:17 UT,
when the filament western end began propagating towards the South.
On August 4, the process of filament elongation and rise accelerated
and the onset of the filament eruption occurred at 03:36~UT and
it fully erupted after 03:48 UT.

\begin{figure}    
  \centerline{\includegraphics[width=0.8\textwidth,clip=]{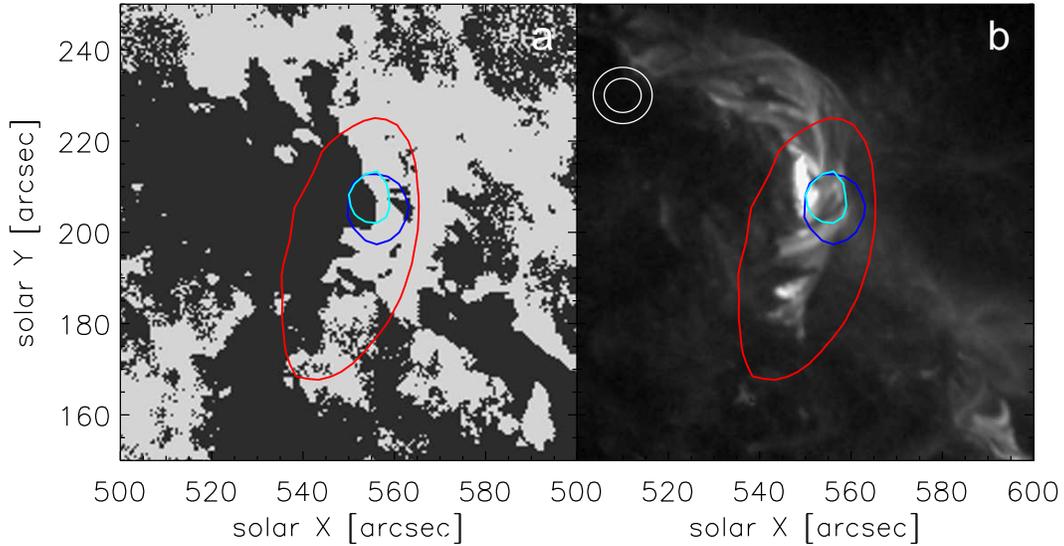}
             }
             \caption{a) Contrast enhanced magnetogram SDO/HMI (03:34:26 UT) and
             half--height contours of microwave sources (03:46 UT) at 5.7~GHz (red), 17~GHz (blue) and 34~GHz (cyan, 03:45:30 UT). Brightness temperature levels of maxima are 2.9 MK, 7.4 MK and 0.6 MK, respectively.
              b) Image of the rising filament and flare emission in line 131{\AA} during Group II of subpeaks at 03:45:35 UT. Circles in top left corner of the
              image show the NORH beam at 17 and 34 GHz.
                     }
  \label{F4-simple}
  \end{figure}

A set of bright EUV loops showed up in the vicinity of the filament southern
tip after  03:36:20 UT. The first microwave enhancement likely associated with the filament rising. Group I of the subpeaks, appeared after 03:44:15 UT.
Figure~\ref{F4-simple} displays the \mw\ sources in the context of the photospheric magnetogram and coronal loops seen in EUV. Figure~\ref{F4-simple}a shows the 50\% contours of the microwave sources at three frequencies during the subpeaks of Group II interval. The SSRT 2D-image, shown in red, is obtained at the end of the interval. 
It is apparent that the 5.7~GHz source is elongated along the neutral line
of the contrast-enhanced magnetogram. The length of the emission region is about 55 arcsec (hereafter, the low-frequency, LF--source). This apparent length is three times larger than the SSRT beam, which was about 20 arcsec; thus, the true LF--source
length can be estimated as roughly 40 arcsec. The apparent width of the source is
comparable with the FWHM of the SSRT beam; thus, the true width does not exceed several arcsec. Therefore, the LF--source can be tentatively identified with
an extended loop.

Unlike the elongated emission source at 5.7 GHz, the sources at 17 GHz
and 34 GHz are not spatially resolved: they have an almost circular shape with
the sizes comparable with the NoRH beam sizes. Given that the source
size was about 10 arcsec at 34 GHz, where the beam width was about 8
arcsec, the true source size can be estimated as $\simeq$ 5~arcsec.
Thus, two spatially distinct sources of the background burst can be identified
in the microwaves: an elongated  LF--source, and a
compact high-frequency HF--source.  As follows from Figure~\ref{F4-simple}b the LF--source covers the southern part of the rising filament.
The HF--source projects on the 131\AA\ image area,
where two small bright ribbons can be distinguished.

\begin{figure}    
  \centerline{\includegraphics[width=1.0\textwidth,clip=]{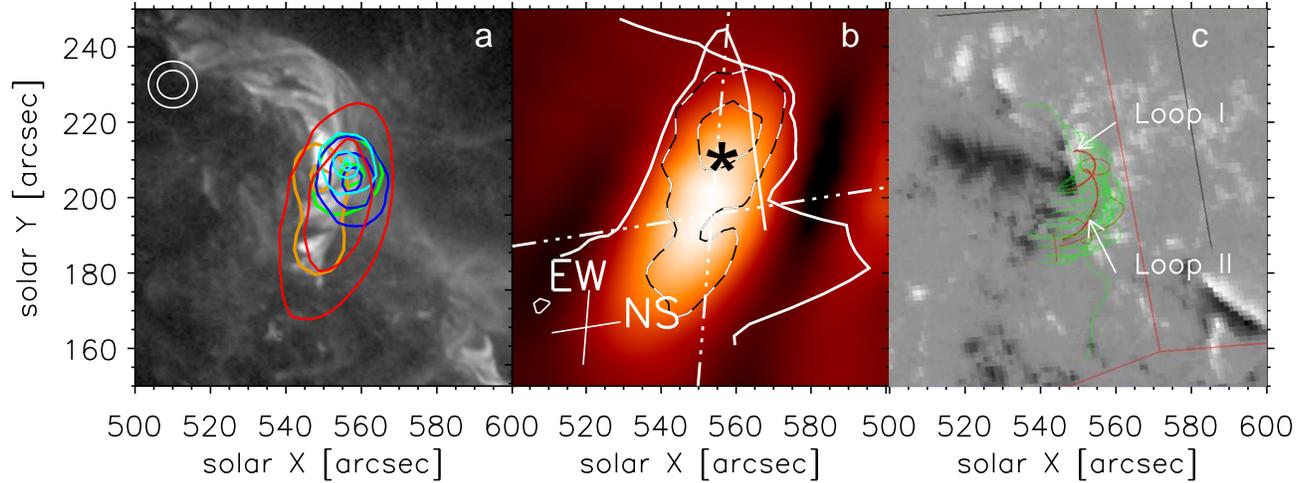}             }
             \caption{a) background is 304~\AA \ image at 03:45:35 UT.
             Red contours are brightness temperature $(0.5, 0.8)\times 2.9$ MK
              at 5.7 GHz at 03:46 UT.  Dark blue contour -- 17~GHz
              (03:45:30 UT, maximum $T_{br}=0.46$ MK); blue -- 34~GHz ($T_{br}=0.08$ MK),  contour levels $(0.3,
              0.7, 0.9)\times T_{br}$; green contour -- 6--12~keV, brown  -- 25--50~keV,
              contour levels (0.5, 0.9) from maximum;
              b)\ background is 5.7~GHz brightness temperature.
              Solid white curves show 1D distributions of brightness during subpeak \# 2.
              These distributions were calculated by subtraction of 1D scans
              at the time: 03:45:20.134 UT and 03:45:18.9 UT. Cross in the lower left corner
              shows the SSRT scanning directions; length of each line is the half
              width of the SSRT beam. The asterisk depicts the
              center of gravity of the narrow-band source during the third group of subpeaks.
              c) \ background: magnetogram SDO/HMI at 03:34:26 UT. Green curves shows the 3D model of field lines, red thick curves marked by arrows depict the lines selected for the microwave source modelling.
                                               }
  \label{F5-simple}
  \end{figure}

In Figure~\ref{F5-simple}a, the HXR sources are shown in two energy bands.
The underlying image shows the map of HeII-line-dominated 304~\AA \ emission at 03:45:35 UT which is the AIA channel most sensitive to the low-temperature plasma of the rising filament.
The brightness center of the  25--50~keV source is located near the
HF--source, but there is an elongated area of the emission along
the LF--source. It's length is about 35 arcsec in the north-south
direction. At 6--12~keV, a presumably thermal, compact source (of about 10 arcsec) projects on the location of
the HF--source.

In the second panel the contours of circularly polarized emission at 5.7~GHz are shown. The region of the polarized emission is extended along the LF--source, whose intensity (Stokes I) is shown as the color image background. The polarization is left-handed everywhere with the degree of polarization below 10\%. The polarization of the 17 GHz source is also left-handed with the degree of polarization about 20\%.

Taken at face value, this combination of the images is consistent
with a single flaring loop elongated in the north-south direction.
The single-loop interpretation, however, suggests a rather unusual
placement of the X-ray sources. Indeed, in such a case, the
low-energy thermal source would be associated with a footpoint of
the loop, while the high-energy nonthermal source would be elongated
along the loop, which is exactly opposite to what is expected and
typically observed. Therefore, we propose that the thermal X-ray
source is associated with compact unresolved flux tubes, while the
HXR emission comes primarily from footpoints of other
loops elongated in the north-south direction. We further propose
that the LF-source is associated with these elongated loops, while
the 17~GHz source is associated with the other (northern compact)
loops, which produces thermal X-ray emission.

It is further important to quantify the spatial configuration of the sources
that generate the short subpeaks of emission and compare it with the
configuration of the source of the background bursts. Unfortunately,
the X-ray imaging available from RHESSI cannot be performed at the
required short time scales. However, microwave imaging can be made
with sufficient cadence to study the subpeak sources. The SSRT scans
(see Section~\ref{S_Instr_mw}) provide the source spatial structure with
the cadence of 14~ms. The NS scans are available over the all length
of Group II, but the EW scans are recorded during the interval with
 subpeaks \# 1-6. Figure~\ref{F5-simple}b shows the scans during
a subpeak marked \# 2 in Figure~\ref{F3-simple}, when the scans were available
from  both SSRT antenna arms. The subpeak component was obtained
as the difference of the scans at the very peak and at the previous
valley. The scans along the flare loops recorded by the NS arm 
show a more elongated source than the EW scans. The scan along 
the flare loops recorded by the NS arm shows
a more elongated source than the EW scan. The size of the subpeak
source is comparable with the size of the background LF--source. The
transverse size slightly exceeds the EW beam, so the actual width of
the subpeak source can be estimated as 10 arcsec. Note that there is
a brightness depression approximately at the center of the NS scan.
The depression length is about 20 arcsec, which is the FWHM of the NS beam. In order for such depression to be observed, the real length of the depression must be $\geq25$ arcsec. Thus, the sources of subpeak emission are adjacent to the loop footpoints.
1D brightness distributions are close for the all subpeaks and position of the depression fluctuated at different subpeaks within 2--8 arcsec along the NS scan, while within less than 4 arcsec along the EW scan. Brightness on the sides of the depression varied; however, they changed synchronously with the accuracy better than 10 ms and a high correlation  between the spatially resolved light curves from the northern and southern regions  with the correlation coefficient of 0.88.
The double structure of the synchronous LF--source brightening is likely a critical observational evidence of particle acceleration in/above the  depression area seen in the NS scans.

The subpeaks at 17 GHz were produced at the HF--source. So, the
position and size of the sources of broadband subpeaks are close to
those  of the background sources at all frequencies. However,
despite similarity of the time profiles, the sizes and locations of
the sources of subpeaks at 5.7 GHz and 17 GHz were significantly
different from each other. These relationships imply a common acceleration region of the nonthermal electrons, but a rather complicated transport of the electrons in involved flaring magnetic flux tubes.

The third group of subpeaks is rather different. The
low-intensity narrowband subpeaks were produced from a compact
source located in the LF--source (Figure~\ref{F5-simple}b). Within the
uncertainty introduced by the SSRT beam size, the estimates yield
the size of the source of a few arcsec. This group of subpeaks does not have any counterpart at other frequencies or HXR energy channels.

\subsection{Spectral properties in X-ray and microwave emission}
\label{S_Obs_Spectral}

\begin{figure}    

 \centerline{\includegraphics[width=0.6\textwidth]{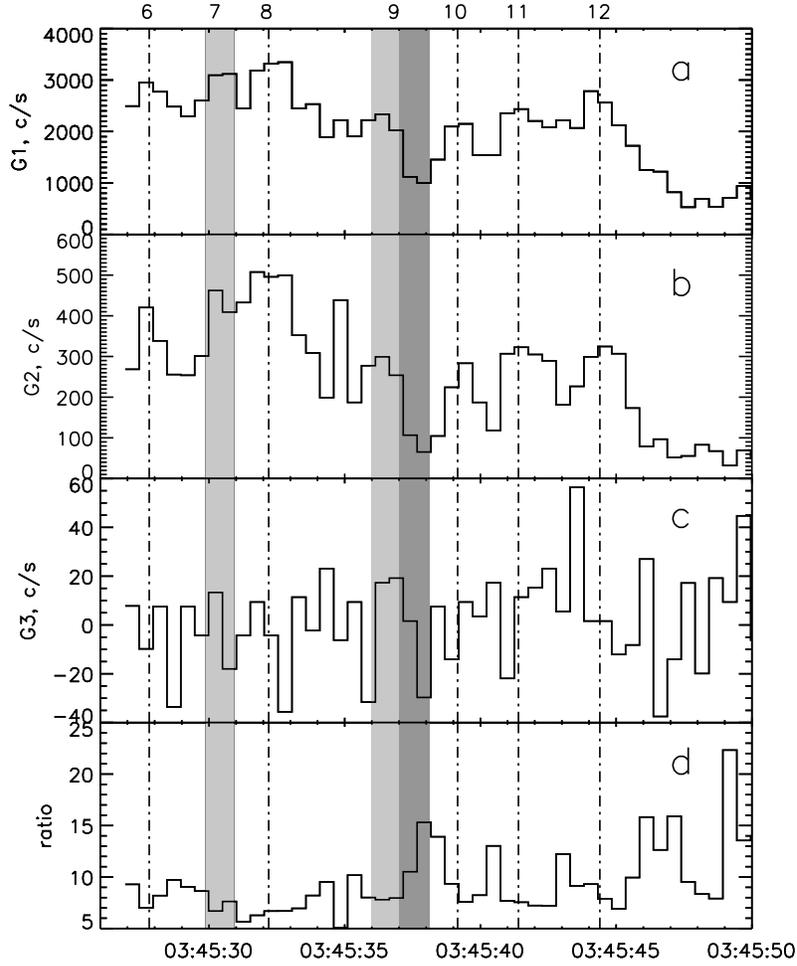}
             }
             \caption{Time profiles of count rate time profiles in the \kw\ energy channels G1 (18--75~keV), G2 (75--308~keV),
             G3 (308--1050~keV), and the ratio G1/G2.
            Temporal resolution is 512 ms.The grey vertical strips show the intervals of the Fermi-GBM   spectrum calculation. 
                     }
  \label{F6-simple}
  \end{figure}

Now, after the temporal and spatial properties of the flaring sources have
been established, we turn to the corresponding spectral properties emphasizing
distinct spectral properties of the subpeaks. Variation of X-ray hardness is quantified using the \kw\ data (Figure~\ref{F6-simple}). The fourth panel of this figure shows the ratio of count rate time profiles in G1 and G2 channels. The ratio decreases during subpeaks, i.e. the subpeak
spectrum is harder compared to the background burst spectrum indicating the soft-hard-soft (SHS) pattern similar to other studies of HXR subsecond variability \citep[e.g.,][]{2016ApJ...822...71F, 2018ApJ...867...84G}.

\begin{figure}    

  \centerline{\includegraphics[width=0.8\textwidth,clip=]{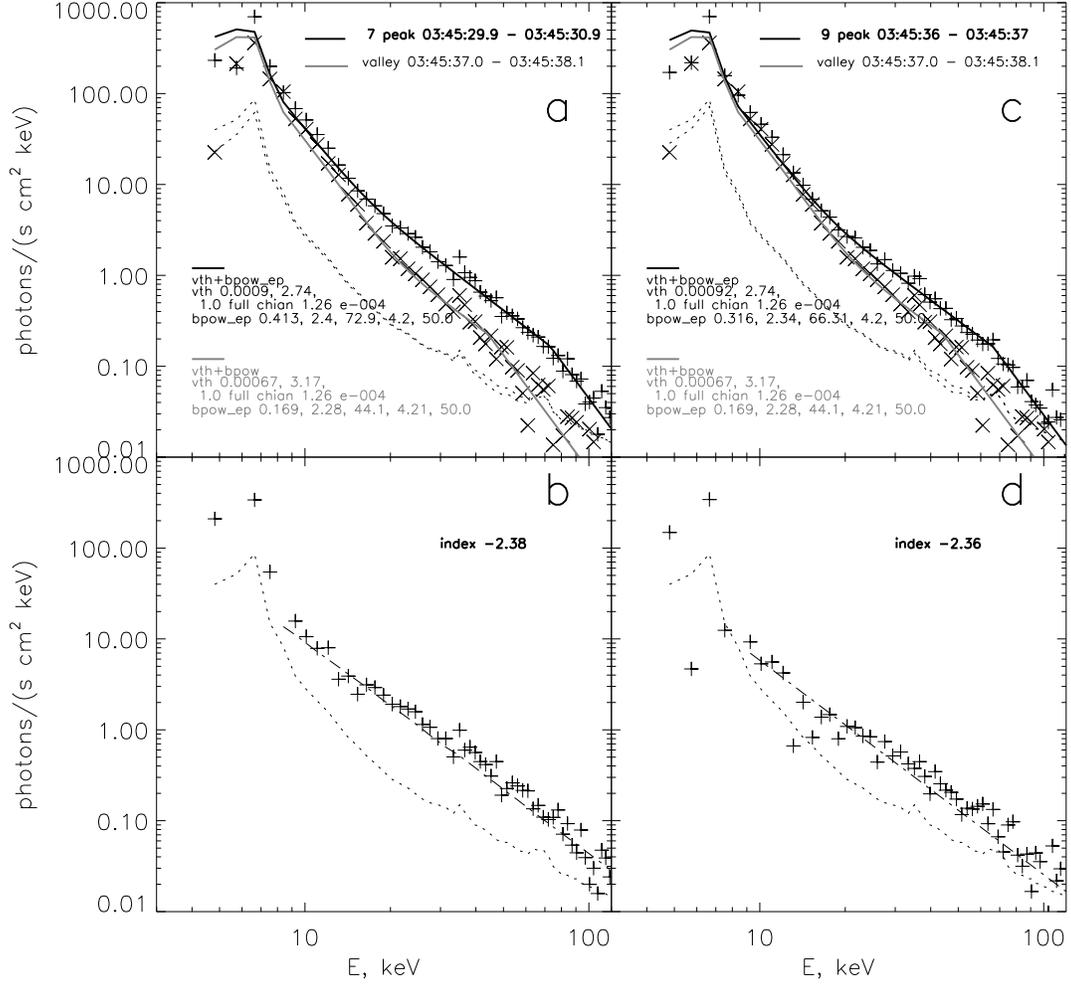}
             }
             \caption{a,c) Spectra of HXR emission recorded by
             the GBM during the 9th and 7th subpeaks (black) and during the
             valley between subpeaks \# 9 and \# 10 (grey). The dotted lines show the background levels. b,d) The differences of these spectra (crosses) and a linear function approximation (dashed).
             The corresponding time intervals are blackened in Figure~\ref{F7-simple}.
             }
  \label{F7-simple}
  \end{figure}

In order to determine properties of the electrons accelerated
during the HXR subpeaks, let us compare the spectra at a subpeak maximum and a valley.
Figures~\ref{F7-simple}a,c show the spectra obtained for subpeaks \# 7 and \# 9 near
the maximum of X-ray signal and for the deep valley between subpeaks \# 9 and \# 10. The spectra were computed from the GBM data using the OSPEX package for the most appropriate 1 s intervals. Subpeak \# 9 occurred directly in front of the valley, while subpeak \# 7 is the brightest one. The intervals with the valley and the subpeak \# 7, 9 are shown by the grey vertical strips in Figure~\ref{F6-simple}. For each time interval we subtracted the time-dependent background from the GBM signals.

The spectra show typical shapes consistent with a thermal core and nonthermal tail,
which can be fitted with a thermal component at low energies and with a broken power-law component at high energy. The temperatures at the valley and at the subpeaks are 30 MK and 37 MK, respectively. The power-law break energies are 44~keV for the valley and 66--73~keV for the peaks,
above which the spectra become softer with indexes $\approx 4.2$.
The power-law indexes under the break energies are 2.3-2.4 and are slightly higher at the time of the subpeaks. Puzzling is the fast change of the plasma temperature suggested by the thermal part of the fit. Indeed, the time scales of the plasma
heating and cooling are typically much longer than the subsecond time
scales of the electron acceleration, so that fast jumps of the plasma
temperature look suspicious. We challenge this straightforward interpretation
of the fit results by computing the difference between
the spectra at the peaks and at the valley (Figure~\ref{F7-simple}b,d). It is seen that their difference in the range between 8~keV and 120~keV is a
power-law function of the photon energy that can be approximated with functions $(2.7\pm0.5)\times10^3 E^{-2.38\pm{0.05}}$ (subpeak \# 7) or $(1.4\pm0.6)\times10^3 E^{-2.36\pm{0.11}}$ (subpeak \# 9), where $E$ is expressed in keV. A similar picture is seen for other peaks, although weaker peaks show larger associated uncertainties.

Thus, our interpretation of the spectrum evolution  is as follows: there is a slowly varying background emission, which consists of a thermal component
 ($T\approx30$~MK) and a broken power-law component, on which a new population
 of freshly accelerated electrons with a single power-law spectrum is superimposed during the peak.

\begin{figure}    

  \centerline{\includegraphics[width=0.8\textwidth,clip=]{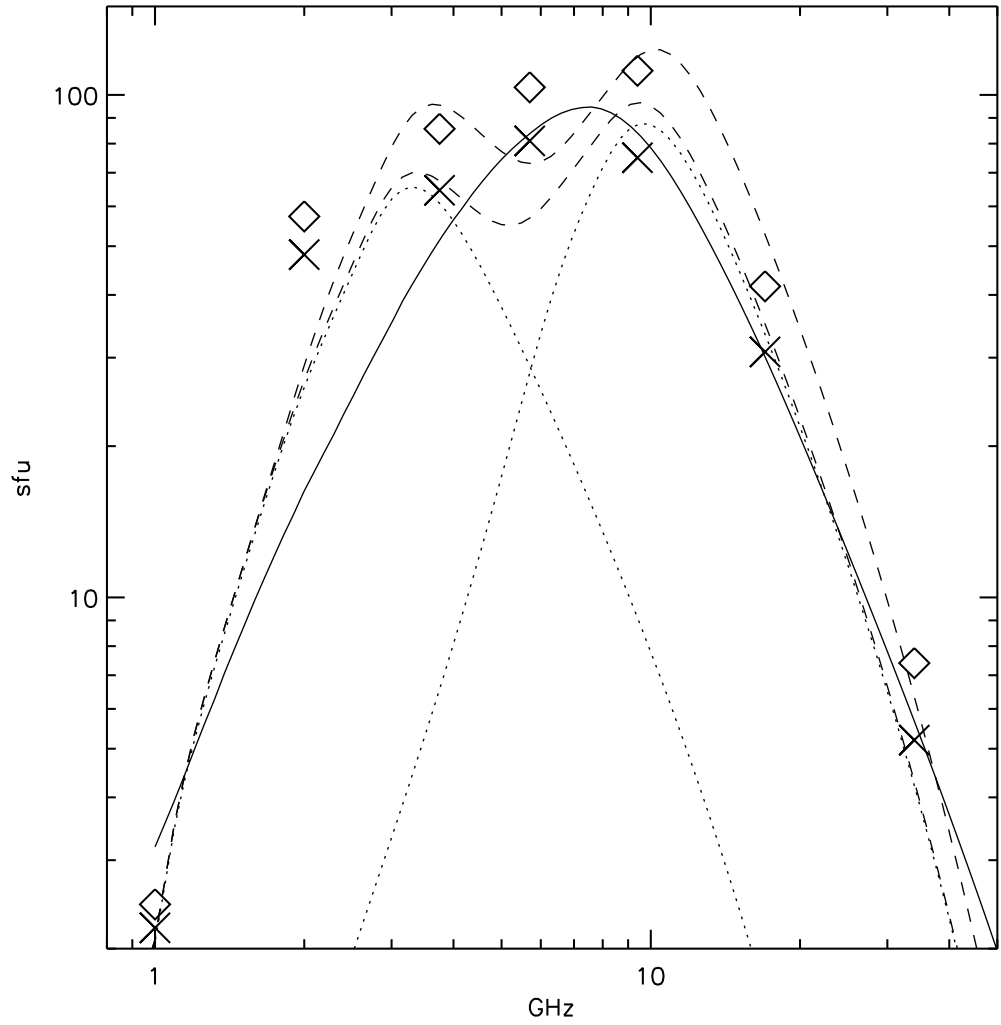}
             }
             \caption{Microwave spectra of subpeak \# 8 maximum (diamonds) and background burst at the valley
             prior subpeak \# 8 (crosses). Solid curve shows the result
             of fitting before subpeak \# 8 using 3D GX Simulator. Results of the GS code fitting are presented by the dashed curves at the subpeak maximum and prior one. Dotted curves show the spectrum contributions of the LF--source and HF--source before subpeak \# 8.
                                  }
  \label{F8-simple}
  \end{figure}

Complementary, evolution of the microwave-emitting electrons can be quantified by time variation 
of the ratio of radio fluxes in the optically thin part of spectra. The
last panel of Figure~\ref{F3-simple} shows the flux ratio at 34
and 17~GHz. It goes up  on average and, thus, the hardness of radio-emitting electrons is slowly increasing on average during interval~II with subpeaks. There is no clear spectral variation pattern associated with subpeaks, which is, perhaps, hidden by a high level of ratio fluctuation. To isolate the contribution to the microwave radiation of electrons injected during the subpeaks, let us consider the spectrum of the largest subpeak in microwaves (\# 8) in comparison with the valley emission in front of it (Figure~\ref{F8-simple}).
The inverted bell shape of the spectra is typical for the gyrosynchrotron emission. Both spectra are flat in the
3.75--9.4~GHz range, which evidences the contribution
to the emission of more than one source or, in a general case, of a
spatial nonuniformity of the source \citep{1994SoPh..153..347L, 2016SoPh..291..445A, 2016ApJ...822...71F}. The spectrum shapes are similar before and at the subpeak, but the difference is weaker at the optically thick part at low frequencies. At the 8th subpeak, for example, the flux intensities are rising up by 30-50\% in the frequency range 3.75--9.4~GHz. We conclude that the overall spectral trends revealed from the \mw\ and HXR data are consistent between each other and consistent with the idea of new injections of the nonthermal electrons during the subpeaks.

\subsection{Summary of Observations and Implications}

The main results for the considered here broadband subpeaks can be summarized as
follows:

\begin{enumerate}

   \item  Configurations of the broadband subpeak and background burst
sources are similar to each other. This configuration consists of two structures.
The first structure (LF--source) is seen at 5.7 GHz emission and
corresponds to loops elongated from the north to the south
for 40 arcsec. The second one is a compact source, which is only
several arcsec in size located at the northern end of the
LF--source. This HF--source  is observed in microwaves at 17 GHz and 34 GHz and in the X-rays. This source is, most likely, associated with a compact flux tube.  The high-energy, 26.7--50.3~keV, X-ray source displays an elongated structure with length of 35 arcsec, superimposed on the LF--source in the picture plane.

   \item  Durations of the subpeaks are  within the range of 0.7--1.8 s.
The short duration and symmetry of the temporal profiles of the subpeaks imply an important role of the scatter-free propagation of the nonthermal electrons in the loops.

   \item  The HXR pulses with photon energy above one hundred keV
lead the lower energy X-ray light curves. Delays in the HXR emission
grow as the energy decreases and reach 70 ms at low energies.
The microwave emission is delayed relative to HXR. The delay at 5.7~GHz is $\sim70$~ms,
while at 17 and 34~GHz is \mytilde 120~ms relative to 102.2--293.5~keV X-rays. 

   \item  There are two sources of subpeak emission at 5.7~GHz which are adjacent to the footpoints of the loop. Their relative brightness varied synchronously, which implies that the electrons are injected from an acceleration region located at/near the looptop towards both footpoints of the loop associated with the LF--source.

   \item   X-ray spectra show SHS behavior during subpeaks. The difference spectrum indicates that such a behavior in the HXRs is caused by pulses of photons with power-law spectrum with indexes about 2.4 above 8~keV.

\end{enumerate}

\section{Discussion}

The observed remarkable similarity between the radio and X-ray light curves at all time scales
from less than a second to several minutes along with revealed spatial relationships between
sources of short peaks and a more gradual background, imply a common origin of the
populations of emitting nonthermal electrons producing both the background burst and the
subpeaks. It is natural to interpret the subpeaks as signatures of fast acceleration of
electrons, a significant fraction of which is then rapidly lost from the source.
To nail down the acceleration mechanism responsible for these short
acceleration episodes and their role in forming the nonthermal component
of the solar flare, we have to quantify the acceleration rate, the spectrum, and the
number of accelerated electrons, as well as the electron transport in the flaring volume.

\subsection{Flare configuration}
 \label{scenario}

To investigate the likely topology of the coronal sources
(loops) involved in the flare, we employ 3D modeling with the GX
Simulator \citep{2015ApJ...799..236N} based on nonlinear force-free field (NLFFF)
reconstruction using the weighted optimization code described and
validated by \citet{2017ApJ...839...30F}. This NLFFF reconstruction is now
a part of the automated model creation pipeline included in the SSW
distribution of the GX Simulator \citep{2018ApJ...853...66N}. The NLFFF reconstruction is initiated with  SDO/HMI vector
magnetogram taken before the flare onset at 03:34:26~UT. The modeling
approach is described in detail in a number of recent publications,
\citep[see, e.g.][and references therein]{2018ApJ...852...32K, 2018ApJ...859...17F}. Specifically, we are looking for magnetic structures consistent with imaging data at the \mw\ and X-ray domains, which form the corresponding flaring flux tubes, and then
populate them with the thermal plasma and nonthermal electrons such
as the synthetic \mw\ and X-ray emissions, computed from the model, match the observations (both
images and spectra).

For our case, the relevant extrapolated magnetic lines forming distinct flux tubes are presented in Fig.~\ref{F5-simple}c. At least two loops are unavoidably needed to reproduce the source morphology (see Section~\ref{S_Obs_Conf}). The two selected loops, whose main parameters are summarized in Table~\ref{T-complex}, permit a reasonable agreement of the calculated \mw\ spectrum with the observed background microwave spectrum (Fig.~\ref{F8-simple}).
Within the model (Fig.~\ref{F5-simple}c), small Loop I is associated with the HF--source, while Loop II is associated with the extended LF--source.

As has already been noted, the nonthermal electron transport in this event is dominated by a scatter-free propagation. This implies that a `trap plus precipitation' (TPP) model \citep{1976MNRAS.176...15M} could be a good approximation for our case. The main feature of this model is a critical pitch-angle of the nonthermal electron population, which demarcates fractions of precipitating and trapped electrons.

\begin{figure}    

  \centerline{\includegraphics[width=0.6\textwidth,clip=]{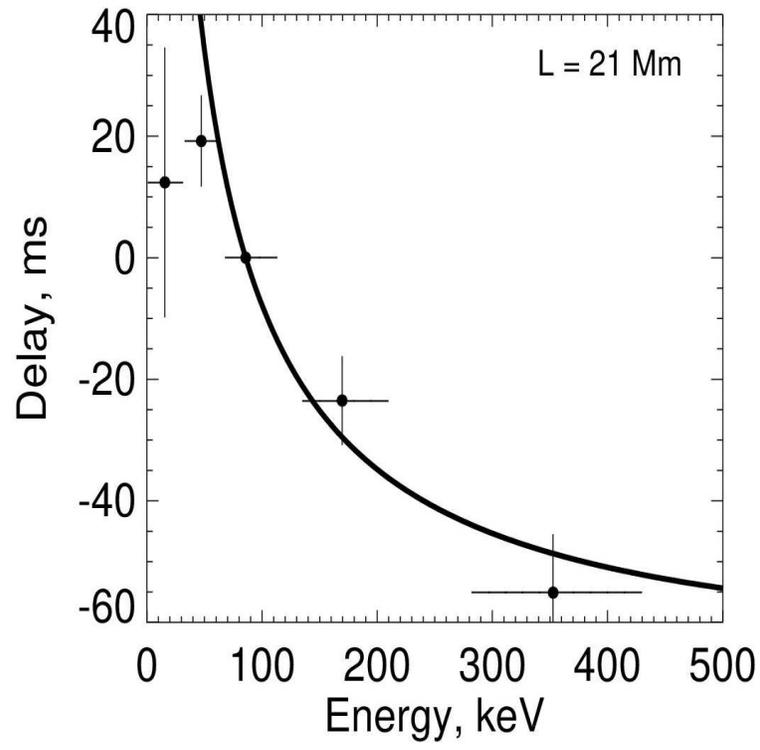}
             }
             \caption{Time delays in various HXR energy channels.
                                  }
  \label{F9-simple}
  \end{figure}

Figure~\ref{F9-simple} shows the relative delays between different HXR channels
as a function of the ``weighted'' energy of the nonthermal electrons that make dominant contribution to a given HXR channel.
These weighted energies (see the fourth column of Table~\ref{T-simple}) were determined based on the algorithm developed by \citet{1996ApJ...464..974A} for a power-law distribution of nonthermal electrons, which we employed for the case of the photon spectral index $\approx2.4$ (see Section~\ref{S_Obs_Spectral} for detail). The measured delays are remarkably consistent with the energy-dependent TOF delays of electrons, which are simultaneously injected into a flaring loop, after propagation over a distance of $L_{TOF} \approx$ 20~Mm (shown in Figure 9 by the solid curve) towards the footpoints; this behavior is similar to a few cases described by (Aschwanden et al. 1996). We note that this ``toy TOF model’’ in its simplest form ignores a number of potentially important physical effects, such as field helical twist of the magnetic field, pitch-angle distribution, evolution of the injected energy spectrum, and potential deviation of the electron transport from the free-streaming mode, any of which can reduce the actual TOF length estimate.

Indeed, the half-length of large loop II is about
11~Mm (see Table~\ref{T-complex}) is in apparent contradiction with the TOF distance (20~Mm) revealed by the delay analysis (Figure~\ref{F9-simple}).  
\cite{1998ApJ...509..911B} demonstrated that a certain type of spectral evolution of the nonthermal electrons precipitating at a footpoint can emulate the same energy-dependent time delays as expected from the TOF propagation even without any propagation (i.e., when the injection occurs just above the footpoint). Such an extreme is clearly inconsistent with our imaging data. In particular, the fact that the microwave subpeaks at two legs of the large loop, mentioned above in Section~\ref{S_Obs_Conf}, unambiguously places the acceleration/injection site at a coronal location at or above the looptop. Having said that, we note that some spectral evolution of the electrons injected at the looptop along with the pitch-angle distribution can still affect the exact delay values, while the field inclination is taken into account within our 3D model. We checked that a SHS evolution of the injected nonthermal electron spectrum, where the spectral index varies by a few percent, while the electrons are beamed along the magnetic field can reduce the TOF distance by a factor of two as needed. However, such a model does not contain any population of nonthermal electrons reflected back to the coronal portion of the flaring loop(s) needed to account the microwave emission. From this, we conclude that an electron beam with a modest angular scatter, rather than an ideal field-aligned beam, is responsible for the observed deviation of the loop half-length from the estimated TOF distance.

Such an angular distribution is capable of reducing their `effective' speed along the magnetic field typically
by a factor of 1.5--3. As we noted above, some angular spread is also required to supply a population of the upward-reflected electrons needed to interpret the microwave data. However, an isotropic angular distribution does not look likely given that it assumes a rather large pitch angle of the majority of streaming electrons and, thus, efficient trapping, which is not observed. Thus, the data favor an anisotropic beam-like distribution of the accelerated electrons with a modest opening angle of $30-40^\circ$.

The mirror ratio $m=3$ in Loop II of our 3D model; thus, the electrons launched downward at the pitch-angle $\vartheta\lesssim 35^\circ$ will not be reflected at the footpoint and, thus, will not contribute to the trapping. The mean parallel velocity $v_\|$  for the electrons launched at the looptop at $\vartheta=  35^\circ$ is $v_\| \approx v \cos\vartheta /2\approx 0.4 v$. Electrons with $\vartheta\gtrsim 35^\circ$ will be reflected and will contribute to the brightness peaks observed at 5.7~GHz from the LF--loop. The observed delay of 70~ms between 5.7~GHz peaks and 102.2-293.5~keV emission peaks,  both produced by presumably  400~keV electrons, places the 5.7~GHz source at a distance $\sim5$~Mm from the mirror points at the bases of loop II.
\Mw\ emission at 17~GHz and 34~GHz delayed by 120~ms requires an additional time for the electrons to be transported from Loop II to Loop I and, perhaps, additional time for trapping in Loop I.

Thus, the time delays are naturally consistent with the identified magnetic configuration of the flare and the TPP model in which the particles are injected at the looptop of Loop II and then propagate to produce the observed variety of the HXR and \mw\ components.

\subsection{Accelerated electron population}

In Section~\ref{S_Obs_Spectral} we determined the HXR spectrum of subpeaks to be a single power-law with the indexes about 2.4. This spectrum is injected during a subpeak in addition to a preexisting population of nonthermal electrons responsible  for a gradual (background) component of the burst.
In the thick target model \citep{1978SoPh...60..137H}, the high energy part of the photon
spectrum is produced by the electron flux of $(0.8 - 3.2)10^{37}E^{-3.4}$~electrons~keV$^{-1}$s$^{-1}$. The total flux of nonthermal electrons above 20~keV, commonly used as a reference energy, can be estimated in this case as $\phi'(>20~{\rm keV})\approx (0.6 - 2.4)10^{34}$s$^{-1}$ during the subpeaks.

Now we address a question if such injections are capable of naturally reproducing properties of the \mw\ subpeaks. To do so, we develop a simplified two-component model guided by imaging data and parameters revealed within 3D modeling.
This simplified modelling employs the fast GS code developed by \citet{2010ApJ...721.1127F} to compute the microwave emission from
two uniform sources with sizes consistent with the SSRT and NoRH maps.
Figure~\ref{F8-simple} and Table~\ref{T-complex}  present the best
simulated spectra of the background microwave emission and associated physical parameters of the sources. These modeling parameters are broadly consistent with the parameters of the flaring volume obtained with the 3D modeling. In particular, the magnetic fields of 250~G and 500~G for the LF- and HF- sources are, respectively, in the ranges of the corresponding magnetic field in the 3D model.

This simplified modelling shows that the increase of emission by 30--50 percent, as observed
during subpeak \# 8, can be explained by roughly doubling the density of non-thermal electrons above 20~keV. The increase of the total number of electrons emitting the microwaves during a subpeak does not exceed $2\times10^{32}$, which is only a few percent of the number of electrons accelerated per pulse. This agrees with the earlier obtained conclusions of relatively inefficient electron trapping in this event and further indicates a reasonably small range of pitch angles of the accelerated electrons.

\subsection{Acceleration mechanism}

The spatially resolved \mw\ data place the acceleration region somewhere in the middle of the extended LF--source.
Given that regardless specific acceleration mechanism, the only force capable of accelerating
charged particles is the electric force, the net electric field (whether regular or stochastic)
has to do the required work $A=\Ef L_{acc}$ on the electron to accelerate it up to half MeV,
where $A\sim 0.5$~MeV and $\Ef$ is the mean electric field along the acceleration path. 

Let us estimate the upper bound of the acceleration length $L_{acc}$ from the timing of the acceleration. Given that the timing of the HXR subpeaks is fully consistent with the timing of precipitation, the acceleration time $\tau_{acc}$ up to high energies must necessarily be much shorter than $\tau_{TOF}\sim85$~ms, thus, necessarily $\tau_{acc}<50$~ms.
The acceleration time $\tau_{acc}<50$~ms implies the acceleration
region size $L_{acc}\sim\left<v\right>\tau_{acc}< 10$~Mm, where $\left<v\right>\approx2\times10^{10}$~cm~s$^{-1}$
is the mean velocity of the electron during the acceleration up to roughly half MeV. For the estimated above upper bound of $L_{acc}\approx10$~Mm, we find the required accelerating electric field $\Ef> A/L_{acc}\sim 0.1$~V/m, which is at least one order of magnitude larger than the typical values of the Dreicer field ($\sim10^{-2}$~V/m).

Super-Dreicer electric fields are both expected \citep{2000mare.book.....P} and implied by the data analysis \citep{2009SoPh..255..107Q, 2016ApJ...822...71F}. However, how exactly these fields drive acceleration of electrons remains very poorly understood. Most likely, a turbulent reconnection environment is formed somewhere at the flare cusp region \citep{2019PPCF...61a4020V}, where multiple current sheets are formed randomly. The reconnection-driven decay of the magnetic field is associated with the electric field according to Faraday equation $\dot{B}=-c\nabla\times{\bf \vE}$, which, in its turn, accelerates electrons to high nonthermal energies. A well-known problem with a traditional acceleration in super-Dreicer electric field in a single current sheet is difficulty of obtaining a power-law spectrum of the nonthermal electrons. In case of the turbulent reconnection with many current sheets involved, the statistics of accelerated electrons may simply follow the statistics of the electric field in the range of available current sheets; thus, if the electric field is distributed over a power-law, the spectrum of the accelerated nonthermal electrons will also follow a power-law \citep{2019PPCF...61a4020V}. In addition, in agreement with observations, such a model can naturally produce a beamed distribution of the accelerated electrons following the preferred direction of the electric field in this collection of the turbulent current sheets, while conventual stochastic acceleration mechanisms tend to produce much more isotropic distributions due to efficient angular scattering of the particles by turbulence.

\section{Conclusions}

The study of the source morphology complemented with the analysis
of timing and spectrum of the subsecond emission peaks,
performed above, has revealed striking properties of the acceleration process.
Indeed, we found that the electrons are accelerated promptly (the acceleration time
is shorter than 50~ms) up to several hundreds of keV, which is difficult
to reconcile with conventual stochastic acceleration models.
These electrons propagate freely along the flux tube, which implies no (or only weak) wave-particle
interaction unlike many other cases \citep[e.g.,][]{2018ApJ...859...17F},
where turbulence was found to play a central role in particle acceleration and transport.
On the other side the observations are broadly consistent with the predictions of models that invoke large, super-Dreicer, electric fields.

\acknowledgements
The work was supported by the Russian Science Foundation (project No. 18-12-00172). G.F. (Discussion) acknowledges support
by NSF grant AGS-1817277 and NASA grants 80NSSC18K0667 and 80NSSC18K1128 to New Jersey Institute of Technology.
A.L. (Konus/Wind data analysis) acknowledges support from RSF grant 17-12-01378.
We are grateful to the teams of the Siberian Solar Radio Telescope, Nobeyama Radio Observatory, SDO, RHESSI, FERMI and \kw  who have provided open access to their data. The experimental data were obtained using the Unique Research Facility Siberian Solar Radio Telescope [http://ckp-rf.ru/usu/73606].

\bibliographystyle{apj}
\bibliography{altref}

\begin{table}
\begin{center}
\caption{The temporal characteristics of the MW and HXR emission}
\label{T-simple}
\begin{tabular}{ccccc}
\hline
Diapason  & Correlation& Observed delay,& Typical\\
& coefficient & ms& energy, keV\\
\hline
4.3--11.6 [keV] & 0.26& 12.4$\pm$ 22.2& 16\\
11.6--26.7 [keV] & 0.88&19.2$\pm$ 7.5&47\\
26.7--50.3 [keV] &  1.00& 0&86\\
50.3--102.2 [keV] & 0.88&-23.5$\pm$ 7.3&170\\
102.2--293.5 [keV]& 0.57&-55.1$\pm$ 9.6 &353\\
5.7[GHz] &0.76&16.7$\pm$ 8.3&200-400 \\
17 [GHz]&0.71&69.5$\pm$ 12.5 & 400-500\\
34 [GHz]&0.62&65.0$\pm$ 17.5& $>$ 700 \\

 \hline
\end{tabular}
\end{center}
\end{table}

\begin{table}
\begin{center}
\caption{Source parameters derived by fitting background microwave spectra}
 \label{T-complex}
\begin{tabular}{lcccc}
\hline
  & \multicolumn{2}{c}{3D GS-Simulator}& \multicolumn{2}{c}{GScode}\\
Source& Loop I & Loop II& HF& LF\\
\hline
Size [Mm]& $15.3\times3\times3$&  $22\times6\times6$ &  $4.3\times4.3\times4.3$& $29\times4.3\times4.3$ \\
Magnetic field (top) [G]&600&300&500&250 \\
Magnetic field (N--footpoint) [G]&1011&916& & \\
Magnetic field (S--footpoint) [G]&-1554&-390& & \\
Plasma density, $n$ [cm$^{-3}$] &$1.5\times10^{10}$&$3.3\times10^{9}$&$10^{10}$&$4\times10^{9}$ \\
Power index, $\delta_{R}$& 2.4&3.3&3.1&2.8\\
Density of non-thermal  &$8.4\times10^{5}$&$1.2\times10^{5}$&$3\times10^{6}$& $5
\times10^{4}$\\
electrons ($>$ 20 keV) [cm$^{-3}$] & & & &\\
Number of
non-thermal &$1.2\times10^{32}$&$9.5\times10^{31}$&$2.3\times10^{32}$&$2.7\times10^{31}$\\
electrons  & & & &\\
 \hline
\end{tabular}
\end{center}
\end{table}

\end{document}